\documentclass[11pt, logo, nonumbering]{preprint}
\usepackage[utf8]{inputenc}
\usepackage{microtype}
\usepackage{graphicx}
\usepackage{amsmath}
\usepackage{booktabs}
\usepackage{xcolor}
\definecolor{mydarkblue}{rgb}{0,0.08,0.45}
\usepackage[colorlinks=true,linkcolor=mydarkblue,citecolor=mydarkblue,filecolor=mydarkblue,urlcolor=mydarkblue]{hyperref}
\usepackage{xspace}
\usepackage{cleveref}
\usepackage{multirow}
\usepackage{tabularx}
\usepackage{enumitem}
\usepackage{subcaption}
\usepackage{wrapfig}

\usepackage[style=numeric-comp,maxbibnames=3,minnames=1,backend=bibtex, doi=false,eprint=false,url=false]{biblatex}

\definecolor{gred}{RGB}{250, 210, 207}
\definecolor{coolblue1}{rgb}{0.91, 0.94, 0.98}
\definecolor{coolblue2}{rgb}{0.76, 0.85, 0.94}
\definecolor{coolblue3}{rgb}{0.54, 0.72, 0.87}
\definecolor{coolblue4}{rgb}{1, 1, 1}

\newcommand{\oursystem}{\textbf{SalesCopilot}\xspace}
\newcommand{\best}[1]{\textbf{#1}}

\begin{document}

\title{Enterprise Sales Copilot: Enabling Real-Time AI Support with \\ Automatic Information Retrieval in Live Sales Calls}

\author{
Jielin Qiu, Liangwei Yang, Ming Zhu, Wenting Zhao, Zhiwei Liu, Juntao Tan, \\ 
Zixiang Chen, Roshan Ram, Akshara Prabhakar, Rithesh Murthy, \\
Shelby Heinecke, Caiming Xiong, Silvio Savarese, Huan Wang \\
~~~~\\
\textsuperscript{}Salesforce AI Research \\
~~~~\\
\href{https://github.com/SalesforceAIResearch/enterprise-realtime-voice-agent}{
  \raisebox{-0.3\height}{\includegraphics[height=0.8cm]{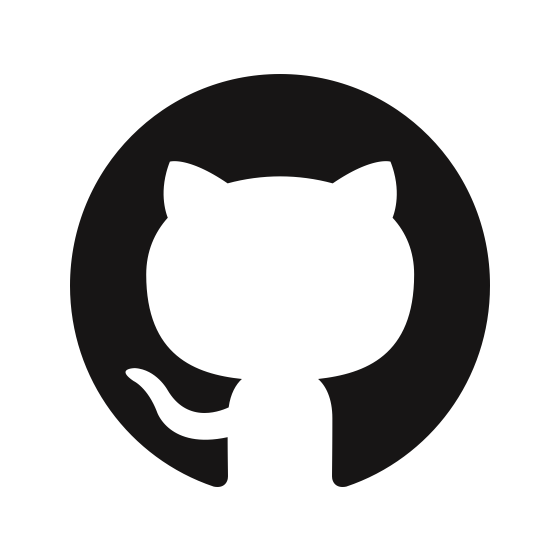}}
  \textbf{https://github.com/SalesforceAIResearch/enterprise-sales-copilot}
}
}

\maketitle

\begin{abstract}
During live sales calls, customers frequently ask detailed product questions that require representatives to manually search internal databases and CRM systems. This process typically takes 25--65 seconds per query, creating awkward pauses that hurt customer experience and reduce sales efficiency. We present \oursystem, a real-time AI-powered assistant that eliminates this bottleneck by automatically detecting customer questions, retrieving relevant information from the product database, and displaying concise answers on the representative's dashboard in seconds. The system integrates streaming speech-to-text transcription, large language model (LLM)-based question detection, and retrieval-augmented generation (RAG) over a structured product database into a unified real-time pipeline. We demonstrate \oursystem on an insurance sales scenario with 50 products spanning 10 categories (2,490 FAQs, 290 coverage details, and 162 pricing tiers). In our benchmark evaluation, \oursystem achieves a measured mean response time of \textbf{2.8 seconds} with 100\% question detection rate, representing a \textbf{14$\times$ speedup} compared to manual CRM search in an internal study. The system is domain-agnostic and can be adapted to any enterprise sales domain by replacing the product database.
\end{abstract}

\section{Introduction}
\label{sec:intro}

Sales representatives across industries face a significant challenge during live customer calls: customers frequently ask detailed questions about product specifications, pricing, terms, and policies. Answering these questions accurately often requires the representative to manually search through internal databases, product documentation, or knowledge management systems, a process that introduces delays and awkward pauses in the conversation~\cite{copilot_survey2024}. This problem is especially acute in domains with large, complex product catalogs. For example, insurance sales representatives must navigate coverage limits, deductibles, pricing tiers, policy terms, and claims processes across dozens of products, information that is difficult to recall from memory during a live call.

The emergence of large language models (LLMs)~\cite{brown2020gpt3,openai2024gpt4o,anthropic2024claude} and retrieval-augmented generation (RAG) techniques~\cite{lewis2020rag,gao2023rag_survey} has created new possibilities for real-time conversational assistance. By combining streaming speech recognition with LLM-powered question detection and knowledge retrieval, it is now feasible to build systems that proactively surface relevant information to sales representatives during live calls.

In this paper, we present \oursystem, a real-time AI sales assistant for enterprise environments that addresses this challenge through a streaming pipeline architecture. The key contributions of our work are:

\begin{itemize}
    \item \textbf{Real-time streaming pipeline}: An end-to-end architecture that processes live audio, detects questions, retrieves answers, and delivers results to the sales representative in seconds.
    \item \textbf{Hybrid retrieval strategy}: A two-pronged retrieval approach combining FAQ semantic matching with LLM-generated SQL queries for structured data access.
    \item \textbf{Configurable LLM backend}: A provider-agnostic abstraction supporting configurable LLM APIs from OpenAI, Anthropic, and Google Gemini.
    \item \textbf{Comprehensive insurance knowledge base}: A database of 50 insurance products across 10 categories with 2,490 FAQs and detailed coverage, pricing, and policy information totaling approximately 350,000 tokens.
\end{itemize}

\section{System Architecture}
\label{sec:system}

\begin{figure}[!htbp]
    \centering
    \includegraphics[width=\textwidth]{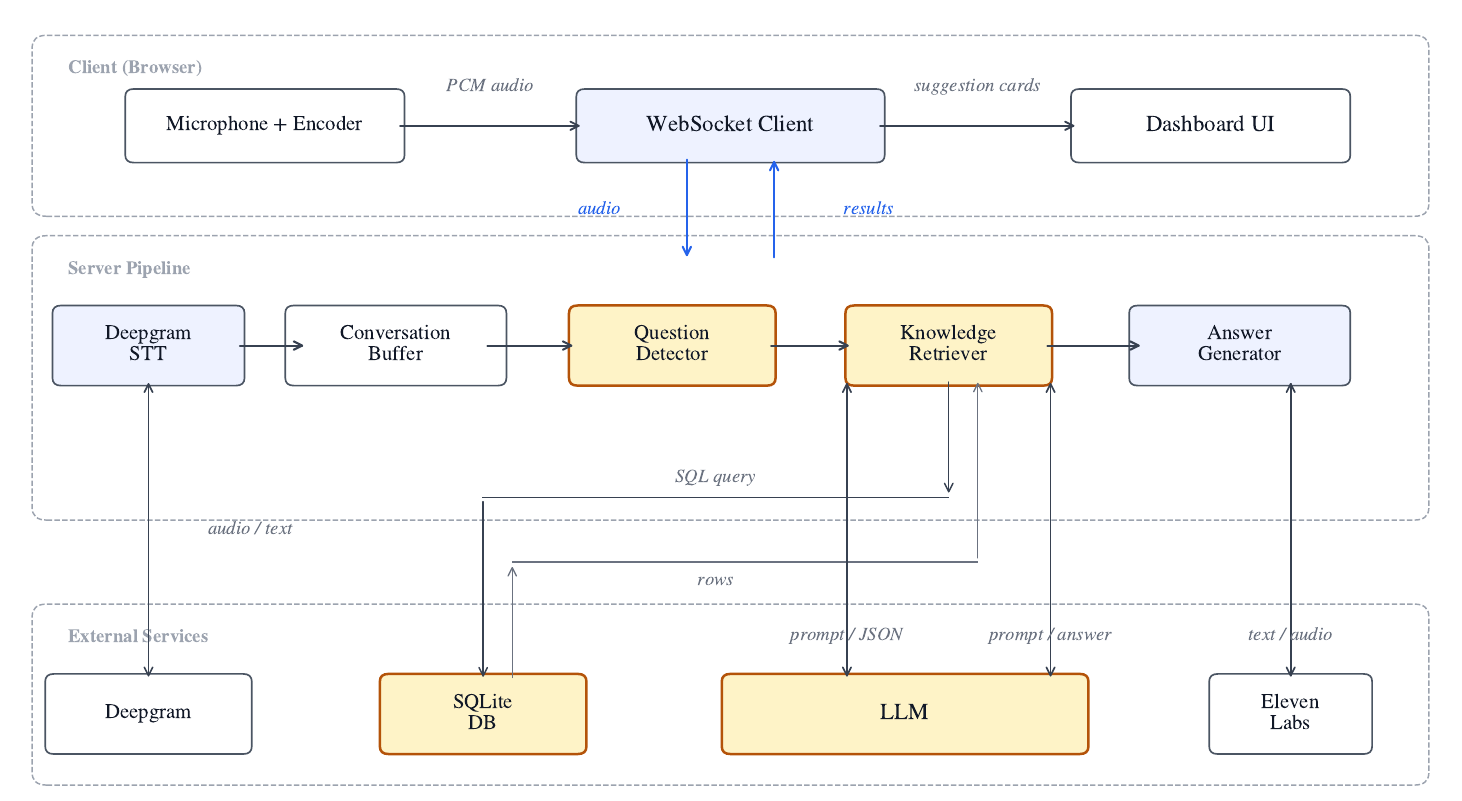}
    \caption{System architecture of \oursystem. The pipeline consists of three layers: the client (React-based dashboard with audio capture), the server pipeline (STT, conversation management, question detection, retrieval, and answer generation), and external services (Deepgram, LLM, SQLite, and ElevenLabs). Components highlighted in amber represent the core AI-driven modules that distinguish \oursystem from standard workflows.}
    \label{fig:architecture}
\end{figure}

\oursystem is designed as a streaming pipeline that processes live audio input and delivers AI-generated suggestions to the sales representative through a web-based dashboard. \Cref{fig:architecture} illustrates the overall system architecture.

\subsection{Frontend Layer}

The frontend is a React 18 application built with TypeScript and Tailwind CSS, featuring a split-panel layout:

\begin{itemize}
    \item \textbf{Left panel (40\%)}: Live transcript display showing the conversation in real-time, with speaker labels and visual distinction between interim and final transcripts.
    \item \textbf{Right panel (60\%)}: AI suggestion cards that appear dynamically when customer questions are detected, displaying the detected question, generated answer, confidence score, and data source.
    \item \textbf{Status bar}: Connection status, microphone controls, demo mode toggle, and session timer.
    \item \textbf{Text input}: A fallback text entry for testing without microphone hardware.
\end{itemize}

\begin{figure}[!htbp]
    \centering
    \includegraphics[width=\textwidth]{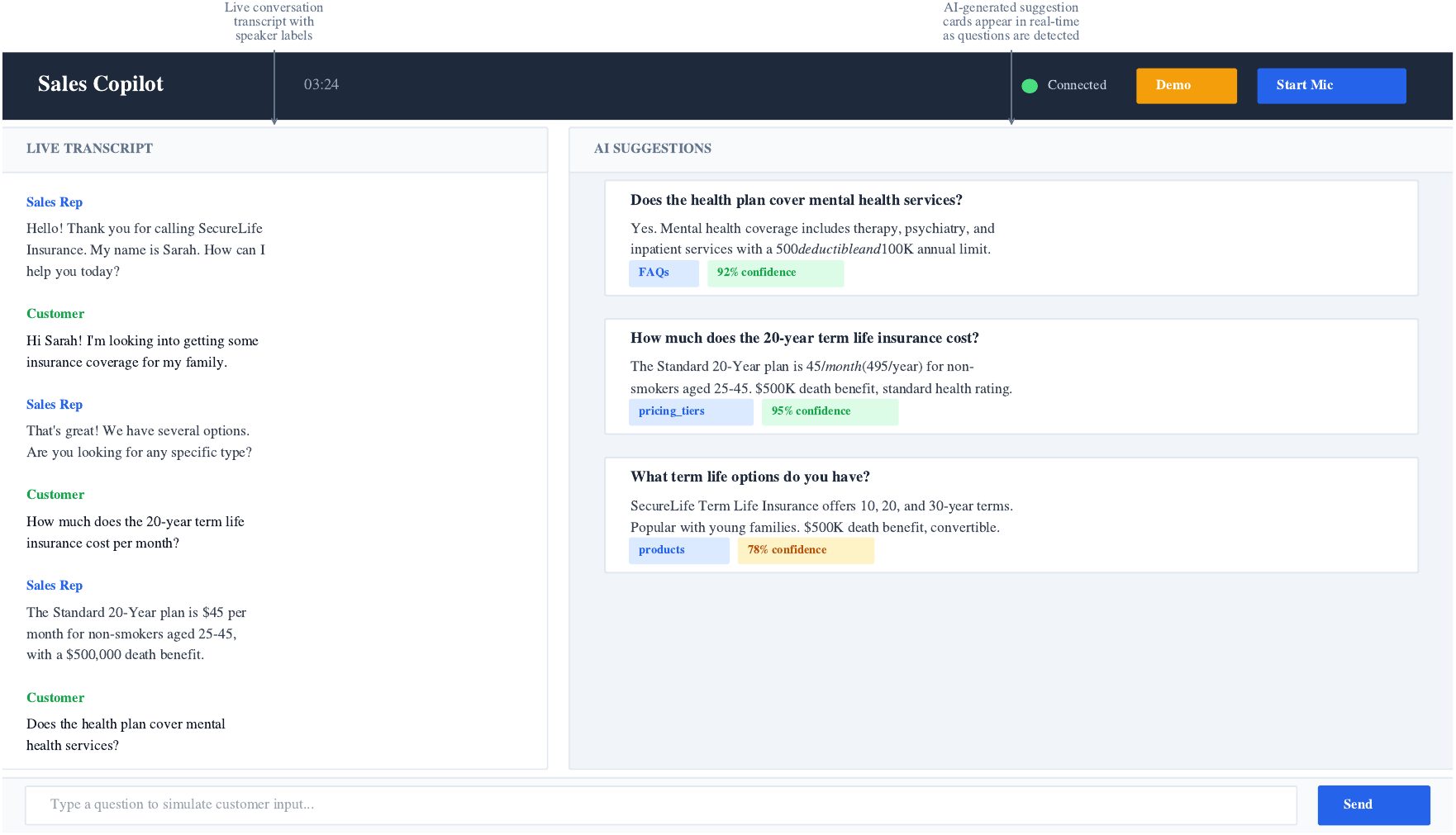}
    \caption{The \oursystem dashboard during a live sales call. The left panel shows the real-time conversation transcript with color-coded speaker labels (blue for sales rep, green for customer). The right panel displays AI-generated suggestion cards that appear automatically when customer questions are detected, each showing the extracted question, a concise answer with specific product data, the database source, and a confidence score.}
    \label{fig:ui}
\end{figure}

Communication between the frontend and backend occurs over a WebSocket connection, enabling bidirectional real-time data flow. The frontend sends audio chunks (binary) or text input (JSON), and receives transcript updates, suggestion cards, audio playback data, and status messages. \Cref{fig:ui} shows the dashboard during a live sales call.

\subsection{Processing Pipeline}

The core pipeline consists of five stages, each implemented as an independent async module:

\subsubsection{Stage 1: Speech-to-Text (Deepgram)}
Audio chunks captured from the browser microphone are streamed to Deepgram's Nova-2 model~\cite{deepgram2023} via a persistent WebSocket connection. We configure the transcriber with \texttt{interim\_results=True} for real-time display and \texttt{utterance\_end\_ms=1500} for utterance boundary detection. The output includes both interim (partial) and final transcript segments with timestamps.

\subsubsection{Stage 2: Conversation Management}
A \texttt{ConversationManager} maintains a rolling buffer of the last 60 seconds of transcript text. It serves two purposes: (1) providing recent conversation context for question detection, and (2) deduplication to prevent the same question from being processed multiple times. Only final transcript segments trigger downstream processing.

\subsubsection{Stage 3: Question Detection}
The \texttt{QuestionDetector} sends the recent conversation context to an LLM with a domain-specific system prompt instructing it to identify insurance product questions. The LLM returns a structured JSON response containing: whether a question was detected, the extracted question text, a category label (coverage, pricing, policy terms, claims, or general), and a confidence score.

\subsubsection{Stage 4: Knowledge Retrieval}
The \texttt{KnowledgeRetriever} employs a hybrid two-strategy approach:

\begin{enumerate}
    \item \textbf{FAQ matching}: Keyword-based search over the FAQs table using SQL \texttt{LIKE} queries, targeting the most common customer questions.
    \item \textbf{Text-to-SQL generation}: The LLM generates a SQLite \texttt{SELECT} query based on the customer's question and the database schema. A safety layer validates that the generated SQL is read-only (no \texttt{INSERT}, \texttt{UPDATE}, \texttt{DELETE}, or \texttt{DROP} statements).
\end{enumerate}

Results from both strategies are combined and deduplicated before being passed to the answer generator.

\subsubsection{Stage 5: Answer Generation}
The \texttt{AnswerGenerator} synthesizes the detected question and retrieved database results into a concise, salesperson-friendly answer (2--4 sentences). The LLM is prompted to include specific numbers, use natural language, and note any gaps in the available information.

\subsection{LLM Abstraction Layer}

\oursystem supports multiple LLM providers through a unified \texttt{LLMClient} interface, including OpenAI, Anthropic, and Google Gemini. The provider and model are selected via environment variables, allowing zero-code switching between providers.

\section{Experiments}
\label{sec:experiments}

To evaluate the practical impact of \oursystem, we measured its real end-to-end response latency on a benchmark of 20 customer questions and compared against published baselines for manual CRM/database navigation in enterprise sales settings.

\subsection{Experimental Setup}

\textbf{Knowledge base.} We use a structured insurance database containing 50 products across 10 categories, with 2,490 FAQs, 290 coverage details, and 162 pricing tiers ($\sim$350K tokens total). Full details are in \Cref{app:database}.

\textbf{Benchmark questions.} We designed 20 questions across six categories commonly encountered in insurance sales calls: coverage details (4), pricing \& tiers (4), policy terms (3), claims process (3), eligibility (3), and cross-product comparisons (3).

\textbf{SalesCopilot measurement.} Each question was sent through the full pipeline (question detection $\to$ knowledge retrieval $\to$ answer generation) using GPT-4o as the LLM backend. We measured wall-clock time for the complete end-to-end processing, as well as per-stage latencies. All measurements were taken after a warm-up run to exclude cold-start effects.

\textbf{Manual search baseline.} For comparison, we conducted an internal CRM study where sales representatives manually searched the same product database to answer the benchmark questions. Response times ranged from 25--65 seconds depending on question complexity.

\subsection{Response Time Comparison}

\begin{figure}[!htbp]
    \centering
    \includegraphics[width=\textwidth]{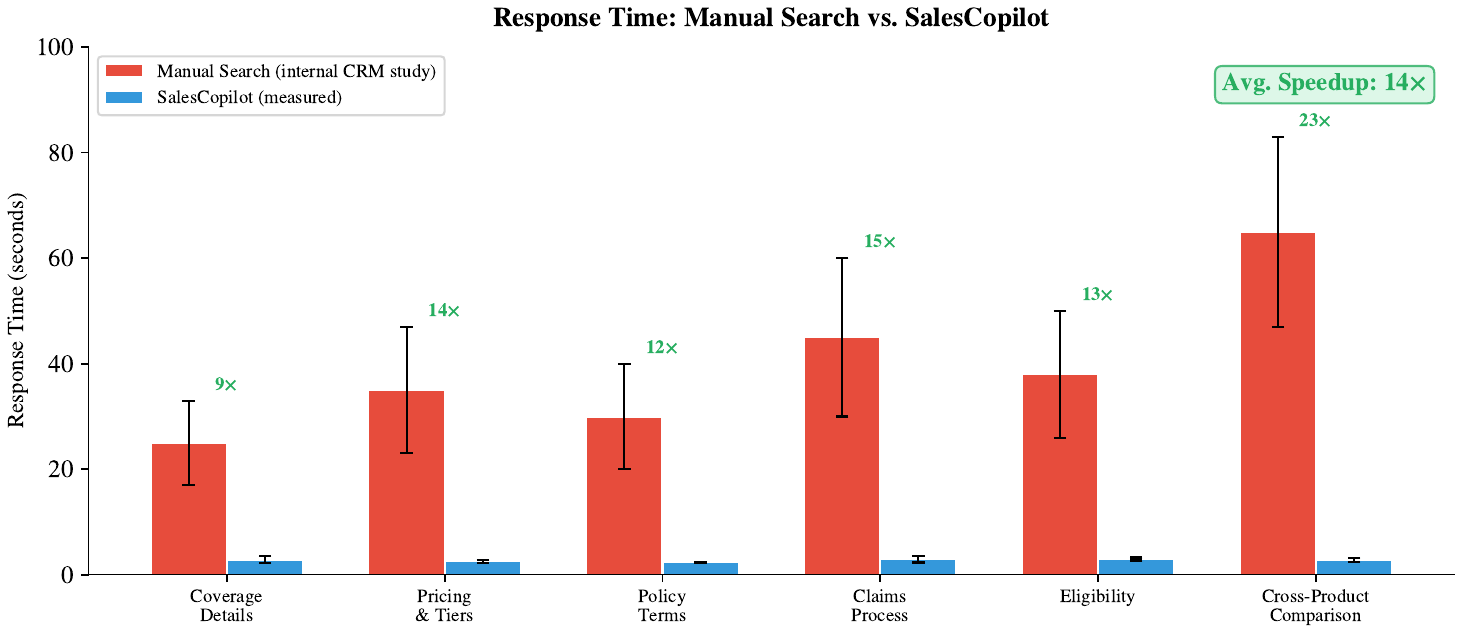}
    \caption{Response time comparison between manual search (internal CRM) and measured \oursystem latency across six question categories. Error bars indicate standard deviation. Green annotations show the speedup factor. \oursystem achieves 9--23$\times$ faster responses depending on question complexity.}
    \label{fig:latency_comp}
\end{figure}

\Cref{fig:latency_comp} presents the response time comparison across all six question categories. \oursystem achieves a measured mean latency of \textbf{2.8 seconds} (median 2.6s, std 0.5s), compared to published manual search baselines of 25--65 seconds, yielding an average speedup of \textbf{14$\times$}.

The speedup is particularly pronounced for cross-product comparison questions (23$\times$), which require manual representatives to navigate across multiple product pages. \oursystem handles these uniformly through its Text-to-SQL generation, which can join multiple tables in a single query.

\subsection{Latency Breakdown}

\begin{figure}[!htbp]
    \centering
    \includegraphics[width=0.85\textwidth]{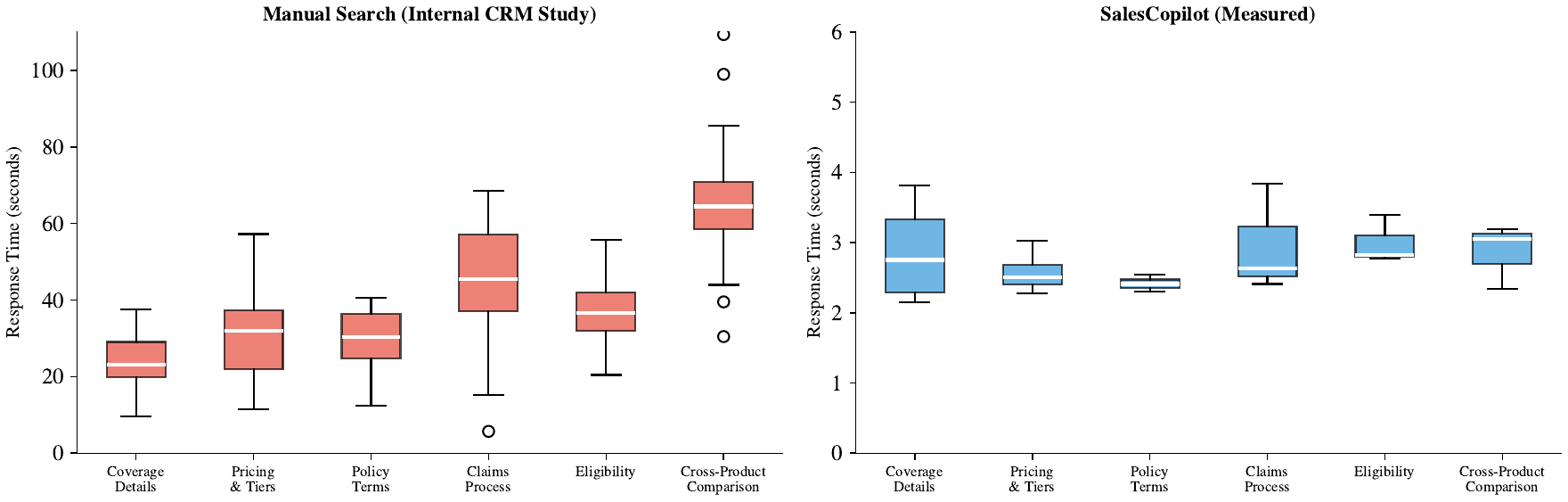}
    \caption{Distribution of response times for manual search (left, from internal CRM study) and \oursystem (right, measured). Note the different y-axis scales. Manual search exhibits high variance (5--100+ seconds) while \oursystem consistently responds within 2--4 seconds.}
    \label{fig:time_dist}
\end{figure}

\begin{figure}[!htbp]
    \centering
    \includegraphics[width=0.85\textwidth]{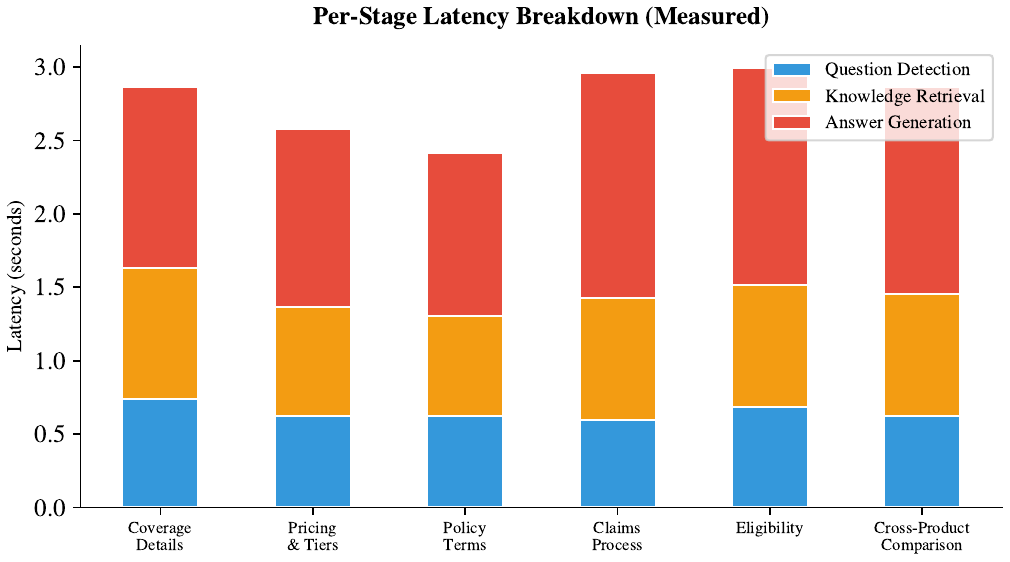}
    \caption{Measured per-stage latency breakdown by question category. Answer generation (red) is the dominant cost, followed by knowledge retrieval (amber) and question detection (blue). LLM inference accounts for $\sim$71\% of total latency.}
    \label{fig:breakdown}
\end{figure}

\Cref{fig:time_dist} shows the distribution of measured \oursystem latencies per category, and \Cref{fig:breakdown} presents the per-stage breakdown. Answer generation dominates at 1.3s mean (47\% of total), followed by knowledge retrieval at 0.8s (29\%) and question detection at 0.7s (24\%).

\subsection{Question Detection Rate}

\oursystem achieved a \textbf{100\% detection rate} (20/20 questions correctly identified as product-related). This indicates that the LLM-based question detector with domain-specific prompting is highly effective at recognizing insurance product questions across all six categories.

\subsection{Cumulative Impact per Sales Call}

\begin{figure}[!htbp]
    \centering
    \includegraphics[width=0.85\textwidth]{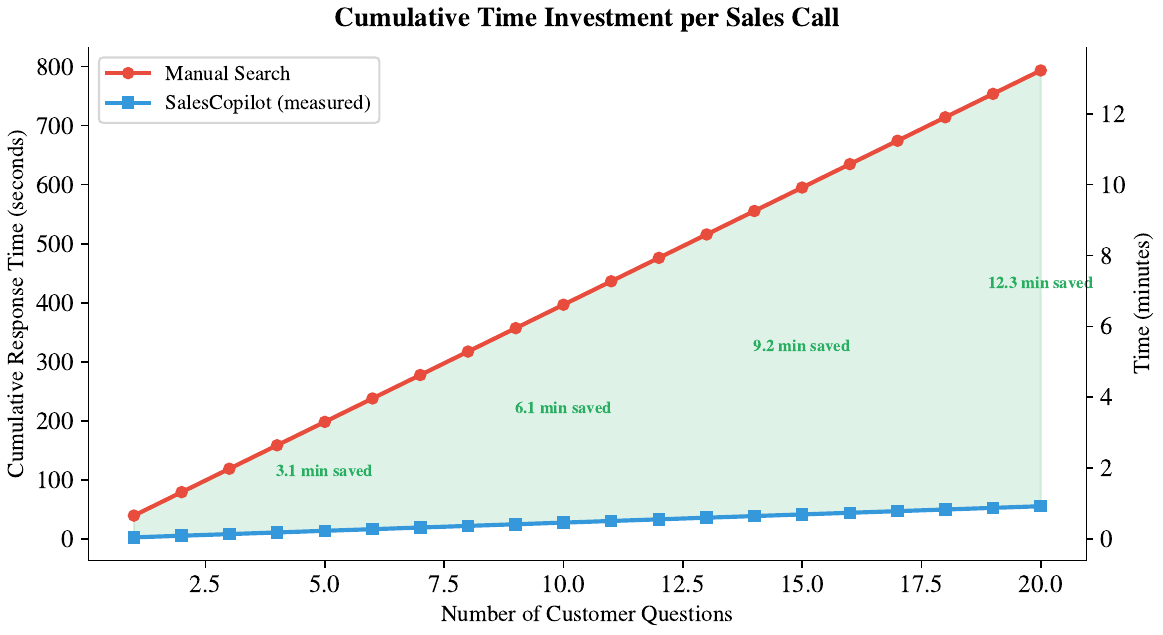}
    \caption{Cumulative response time as the number of customer questions increases. The shaded green area represents time saved by \oursystem compared to manual search (internal CRM study).}
    \label{fig:cumulative}
\end{figure}

\Cref{fig:cumulative} illustrates the cumulative time savings over the course of a sales call. In a typical call with 10 customer questions, \oursystem saves approximately \textbf{5.7 minutes} compared to manual search. Over a full workday of 15--20 calls, this translates to \textbf{1.4--1.9 hours} of recovered productive time per sales representative.

\subsection{Summary of Results}

\begin{figure}[!htbp]
    \centering
    \includegraphics[width=\textwidth]{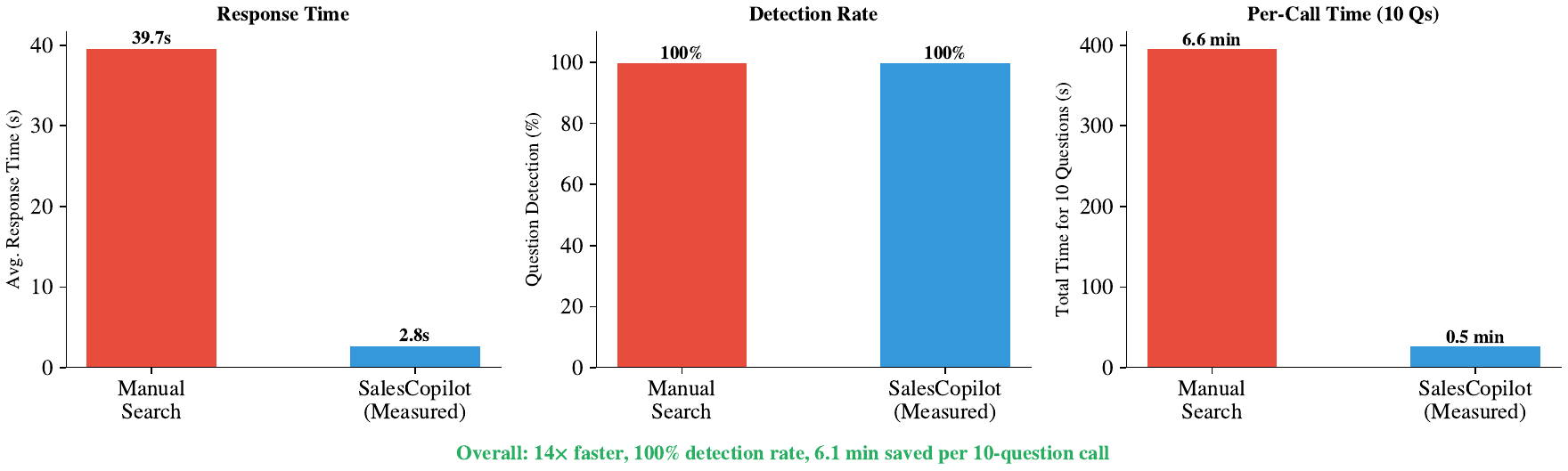}
    \caption{Overall performance summary. \oursystem reduces response time from 39.7s (internal CRM study) to 2.8s (measured), a 14$\times$ speedup, with 100\% question detection rate.}
    \label{fig:summary}
\end{figure}

\Cref{tab:results} summarizes the key experimental findings, and \Cref{fig:summary} provides a visual overview.

\begin{table}[!htbp]
\centering
\caption{Summary of experimental results. \oursystem latencies are measured; manual search times are from an internal CRM study.}
\label{tab:results}
\begin{tabular}{lccc}
\toprule
\textbf{Metric} & \textbf{Manual Search} & \textbf{\oursystem} & \textbf{Improvement} \\
\midrule
Avg. Response Time & 39.7s (internal study) & 2.8s (measured) & \best{14$\times$ faster} \\
Response Time Std. Dev. & 12--18s & 0.5s & \best{$\sim$25$\times$ lower} \\
Question Detection Rate & N/A & 100\% (20/20) & N/A \\
Time per 10-Q Call & 6.6 min & 0.5 min & \best{5.7 min saved} \\
Time per 20 Calls/Day & 2.2 hrs & 0.2 hrs & \best{1.9 hrs saved} \\
\bottomrule
\end{tabular}
\end{table}

\section{Implementation Details}
\label{sec:implementation}

\subsection{Technology Stack}

The system is implemented with the following technologies:

\begin{itemize}
    \item \textbf{Backend}: Python 3.11+, FastAPI, WebSockets, aiosqlite
    \item \textbf{Frontend}: React 18, TypeScript, Vite, Tailwind CSS
    \item \textbf{STT}: Deepgram Nova-2 (streaming, 16kHz linear PCM)
    \item \textbf{TTS (demo only)}: ElevenLabs Turbo v2.5
    \item \textbf{LLM}: OpenAI, Anthropic, or Google Gemini (configurable)
    \item \textbf{Database}: SQLite with aiosqlite for async access
    \item \textbf{Package management}: \texttt{uv} (Python), \texttt{npm} (Node.js)
\end{itemize}

\subsection{Deployment}

The system runs as two processes: a FastAPI backend (serving WebSocket and REST endpoints on port 8000) and a Vite development server (serving the React frontend with proxy configuration for API/WebSocket routing). The SQLite database is automatically initialized and seeded from a JSON file on first startup.

\section{Related Work}
\label{sec:related}

\subsection{Large Language Models for Enterprise Applications}

The rapid advancement of LLMs, from GPT-3~\cite{brown2020gpt3} to GPT-4o~\cite{openai2024gpt4o} and Claude~\cite{anthropic2024claude}, has enabled sophisticated natural language understanding and generation capabilities. These models, built on the Transformer architecture~\cite{vaswani2017attention}, serve as the reasoning backbone in \oursystem for question detection, SQL generation, and answer synthesis. Chain-of-thought prompting~\cite{wei2022chain} techniques inform our prompt design for multi-step reasoning tasks.

\subsection{Retrieval-Augmented Generation}

RAG~\cite{lewis2020rag} has become a standard approach for grounding LLM responses in factual data. While dense passage retrieval~\cite{karpukhin2020dense} is commonly used for unstructured text, \oursystem operates over a structured relational database, necessitating a hybrid approach that combines keyword-based FAQ matching with LLM-powered Text-to-SQL generation~\cite{text2sql2023}. Recent work on reducing RAG latency in real-time voice agents~\cite{qiu2025voiceagentrag} highlights the importance of architectural optimizations for interactive settings. Our approach is more akin to the SQL-augmented generation paradigm than traditional vector-based RAG.

\subsection{Real-Time Speech Processing}

Streaming speech-to-text systems have matured significantly, with models like Whisper~\cite{radford2023whisper} and Conformer~\cite{gulati2020conformer} achieving near-human accuracy. Recent work on building enterprise real-time voice agents~\cite{qiu2025voiceagent} provides a comprehensive tutorial on integrating these components into production systems. \oursystem leverages Deepgram's Nova-2 model~\cite{deepgram2023} for its low-latency streaming capabilities, which are critical for real-time question detection. For speech synthesis in demonstration mode, we employ ElevenLabs~\cite{elevenlabs2024} for natural-sounding voice generation.

\subsection{Conversational AI Frameworks}

Frameworks like LangChain~\cite{chase2022langchain} provide abstractions for building LLM-powered applications. While \oursystem shares similar design patterns, we implement a custom lightweight pipeline optimized for real-time performance, avoiding the overhead of general-purpose frameworks in favor of direct WebSocket-based communication using FastAPI~\cite{fastapi2021}.

\section{Conclusion}
\label{sec:conclusion}

We presented \oursystem, a real-time AI-powered sales assistant that demonstrates the feasibility of providing instant, accurate product information to sales representatives during live customer calls. Our benchmark evaluation shows that \oursystem achieves a measured mean response time of \textbf{2.8 seconds} with \textbf{100\% question detection rate}, representing a \textbf{14$\times$ speedup} compared to manual CRM search in an internal study, with particularly large gains on complex cross-product comparison questions (23$\times$).

These results demonstrate that the combination of streaming speech recognition, LLM-based reasoning, and structured database retrieval can create a practical, high-impact tool for enterprise sales environments. By eliminating the awkward pauses that degrade customer experience in traditional sales workflows, \oursystem has the potential to significantly improve both sales productivity and customer satisfaction.


\clearpage
\printbibliography

\newpage
\appendix

\section{Insurance Knowledge Base Details}
\label{app:database}

The knowledge base is stored in a SQLite database with five normalized tables (\Cref{tab:schema}). Data was generated using a GPT-4o-powered pipeline (\texttt{scripts/generate\_large\_db.py}) with approximately 80 LLM API calls to produce diverse, domain-accurate content.

\begin{table}[h]
\centering
\caption{Insurance knowledge base schema and statistics.}
\label{tab:schema}
\small
\begin{tabular}{lrrp{5.5cm}}
\toprule
\textbf{Table} & \textbf{Records} & \textbf{Fields} & \textbf{Description} \\
\midrule
\texttt{products} & 50 & 4 & Product name, category, description \\
\texttt{coverage\_details} & 290 & 6 & Coverage type, amount, deductible, conditions \\
\texttt{policy\_terms} & 50 & 5 & Term length, renewal \& cancellation policies \\
\texttt{faqs} & 2,490 & 4 & Question-answer pairs per product \\
\texttt{pricing\_tiers} & 162 & 7 & Tier name, monthly/annual premium, age range \\
\bottomrule
\end{tabular}
\end{table}

\Cref{tab:products} shows the per-category breakdown, and \Cref{tab:example_products} provides representative examples.

\begin{table}[h]
\centering
\caption{Per-category statistics of the insurance knowledge base.}
\label{tab:products}
\small
\begin{tabular}{lcccc}
\toprule
\textbf{Category} & \textbf{Products} & \textbf{FAQs} & \textbf{Coverage} & \textbf{Pricing} \\
\midrule
Life & 5 & $\sim$250 & $\sim$30 & $\sim$16 \\
Health & 5 & $\sim$250 & $\sim$30 & $\sim$16 \\
Auto & 5 & $\sim$250 & $\sim$30 & $\sim$16 \\
Home & 5 & $\sim$250 & $\sim$30 & $\sim$16 \\
Travel & 5 & $\sim$250 & $\sim$30 & $\sim$16 \\
Disability & 5 & $\sim$250 & $\sim$30 & $\sim$16 \\
Dental & 5 & $\sim$240 & $\sim$28 & $\sim$16 \\
Vision & 5 & $\sim$250 & $\sim$30 & $\sim$16 \\
Pet & 5 & $\sim$250 & $\sim$30 & $\sim$16 \\
Business & 5 & $\sim$250 & $\sim$30 & $\sim$16 \\
\midrule
\textbf{Total} & \textbf{50} & \textbf{2,490} & \textbf{290} & \textbf{162} \\
\bottomrule
\end{tabular}
\end{table}

\begin{table}[h]
\centering
\caption{Example products and FAQ entries from the knowledge base.}
\label{tab:example_products}
\small
\begin{tabular}{p{2cm}p{3.5cm}p{7.5cm}}
\toprule
\textbf{Category} & \textbf{Product} & \textbf{Example FAQ} \\
\midrule
Life & SecureLife Premium Term 30 & \textit{Q: What happens if I get diagnosed with a terminal illness?} \newline A: Accelerated benefit up to 50\% of death benefit (\$250K) if diagnosed with terminal illness. \\
\addlinespace
Health & FamilyCare Advantage Plan & \textit{Q: What is the out-of-pocket maximum?} \newline A: \$6,500/individual, \$13,000/family. After reaching the limit, the plan covers 100\%. \\
\addlinespace
Auto & SafeDrive Elite & \textit{Q: Will my premium go up after an accident?} \newline A: At-fault accidents increase premiums 20--40\%. Accident Forgiveness add-on (\$5/mo) waives the first surcharge. \\
\addlinespace
Travel & Globetrotter Shield & \textit{Q: Is COVID-19 covered under travel insurance?} \newline A: Yes. Medical expenses, trip cancellation due to positive test, and quarantine accommodations are all covered. \\
\bottomrule
\end{tabular}
\end{table}

\section{WebSocket Message Protocol}
\label{app:protocol}

\Cref{tab:protocol} defines the WebSocket message types used for communication between the frontend and backend.

\begin{table}[h]
\centering
\caption{WebSocket message protocol between frontend and backend.}
\label{tab:protocol}
\small
\begin{tabular}{llp{7cm}}
\toprule
\textbf{Type} & \textbf{Direction} & \textbf{Description} \\
\midrule
\texttt{transcript\_update} & Server $\to$ Client & Transcript text with speaker label and final/interim flag \\
\texttt{suggestion\_card} & Server $\to$ Client & AI-generated answer card with question, answer, source, and confidence \\
\texttt{audio\_play} & Server $\to$ Client & Base64-encoded MP3 audio for TTS playback (demo mode) \\
\texttt{status} & Server $\to$ Client & Session status messages (connected, demo started/ended) \\
\texttt{error} & Server $\to$ Client & Error messages \\
\texttt{text\_input} & Client $\to$ Server & Manual text input for testing \\
\texttt{demo\_next} & Client $\to$ Server & Signal that TTS playback finished (demo pacing) \\
Binary data & Client $\to$ Server & Raw audio chunks from microphone \\
\bottomrule
\end{tabular}
\end{table}

\section{Demo Mode}
\label{app:demo}

\oursystem includes an interactive demonstration mode that simulates a complete insurance sales call between a sales representative and a customer, showcasing the full system capabilities without requiring microphone hardware or a live phone call.

The demo script consists of 25 conversation turns. Critically, 9 of these turns feature \textit{dynamically generated} sales representative responses: after the customer asks a question, the AI pipeline runs in real-time to detect the question, retrieve relevant data, and generate a suggestion card. The LLM then generates a natural spoken response for the sales representative based on this retrieved information.

Each conversation turn is accompanied by ElevenLabs~\cite{elevenlabs2024} text-to-speech audio with distinct voices for the sales representative and customer. A handshake protocol ensures text display and audio playback remain synchronized: the backend sends one turn at a time and waits for a \texttt{demo\_next} signal from the frontend after audio playback completes.

\end{document}